# Discovery of a Weyl Fermion Semimetal and Topological Fermi Arcs

**Authors:** Su-Yang Xu†,[1,2] Ilya Belopolski†,[1] Nasser Alidoust†,[1,2] Madhab Neupane†,[1,3] Guang Bian,[1] Chenglong Zhang,[4] Raman Sankar,[5] Guoqing Chang,[6,7] Zhujun Yuan,[4] Chi-Cheng Lee,[6,7] Shin-Ming Huang,[6,7] Hao Zheng,[1] Jie Ma,[8] Daniel S. Sanchez,[1] BaoKai Wang,[6,7,9] Arun Bansil,[9] Fangcheng Chou,[5] Pavel P. Shibayev,[1,10] Hsin Lin,[6,7] Shuang Jia,[4,11] and M. Zahid Hasan*[1,2]

**Abstract**: A Weyl semimetal is a crystal which hosts Weyl fermions as emergent quasiparticles and admits a topological classification that protects Fermi arc surface states on the boundary of a bulk sample. This unusual electronic structure has deep analogies with particle physics and leads to unique topological properties. We report the experimental discovery of the first Weyl semimetal, TaAs. Using photoemission spectroscopy, we directly observe Fermi arcs on the surface, as well as the Weyl fermion cones and Weyl nodes in the bulk of TaAs single crystals. We find that Fermi arcs terminate on the Weyl nodes, consistent with their topological character. Our work opens the field for the experimental study of Weyl fermions in physics and materials science.

NoteAdded: *This experimental discovery (Science 2015) is based on our earlier 2014 theoretical discovery/prediction reported at [Huang et al., Nature Commun. **6**:7373 (2015) **submitted (Nov. 2014)**. http://www.nature.com/ncomms/2015/150612/ncomms8373/full/ncomms8373.html.*

*This is the first theoretical prediction of TaAs being a Weyl semimetal with Fermi arcs. All other theory papers were submitted at least 6 weeks after our theoretical prediction of TaAs, see http://www.nature.com/ncomms/2015/150612/ncomms8373/full/ncomms8373.html.]*

**Affiliations:**

[1]Laboratory for Topological Quantum Matter and Spectroscopy (B7), Department of Physics, Princeton University, Princeton, New Jersey 08544, USA.

[2]Princeton Center for Complex Materials, Princeton Institute for Science and Technology of Materials, Princeton University, Princeton, New Jersey 08544, USA.

[3]Condensed Matter and Magnet Science Group, Los Alamos National Laboratory, Los Alamos, New Mexico 87545, USA.

[4]International Center for Quantum Materials, School of Physics, Peking University, China.

[5]Center for Condensed Matter Sciences, National Taiwan University, Taipei 10617, Taiwan.

[6]Centre for Advanced 2D Materials and Graphene Research Centre National University of Singapore, 6 Science Drive 2, Singapore 117546.

[7]Department of Physics, National University of Singapore, 2 Science Drive 3, Singapore 117542

[8]Quantum Condensed Matter Division, Oak Ridge National Laboratory, Oak Ridge, Tennessee 37831, USA.

[9]Department of Physics, Northeastern University, Boston, Massachusetts 02115, USA.

[10]Princeton Institute for Science and Technology of Materials, Princeton University, Princeton, New Jersey 08544, USA.

[11]Collaborative Innovation Center of Quantum Matter, Beijing, 100871, China.

† These authors contributed equally to this work.

* Corresponding author. E-mail: mzhasan@Princeton.edu

Weyl fermions have long been known in quantum field theory, but have not been observed as a fundamental particle in nature (*1-3*). Recently, it was understood that a Weyl fermion can emerge as a quasiparticle in certain crystals, Weyl fermion semimetals (*1-22*). Despite being a gapless metal, a Weyl semimetal is characterized by topological invariants, broadening the classification of topological phases of matter beyond insulators. Specifically, Weyl fermions at zero energy correspond to points of bulk band degeneracy, Weyl nodes, which are associated with a chiral charge that protects gapless surface states on the boundary of a bulk sample. These surface states take the form of Fermi arcs connecting the projection of bulk Weyl nodes in the surface Brillouin zone (BZ) (*6*). A band structure like the Fermi arc surface states would violate basic band theory in an isolated two-dimensional system and can only arise on the boundary of a three-dimensional sample, providing a dramatic example of the bulk-boundary correspondence in a topological phase. In contrast to topological insulators where only the surface states are interesting (*21*, *22*), a Weyl semimetal features unusual band structure in the bulk and on the surface. The Weyl fermions in the bulk are predicted to provide a condensed matter realization of the chiral anomaly, giving rise to a negative magnetoresistance under parallel electric and magnetic fields, unusual optical conductivity, non-local transport and local non-conservation of ordinary current (*5*, *12-16*). At the same time, the Fermi arc surface states are predicted to show novel quantum oscillations in magneto-transport, as well as unusual quantum interference effects in tunneling spectroscopy (*17-19*). The prospect of the realization of these unusual phenomena has inspired much experimental and theoretical work. (*1-22*).

Here we report the experimental realization of a Weyl semimetal in a single crystalline material tantalum arsenide, TaAs. Utilizing the combination of the vacuum ultraviolet (low-photon-energy) and soft X-ray (SX) angle-resolved photoemission spectroscopy (ARPES), we systematically and differentially study the surface and bulk electronic structure of TaAs. Our ultraviolet (low-photon-energy) ARPES measurements, which are highly surface sensitive, demonstrate the existence of the Fermi arc surface states, consistent with our band calculations presented here. Moreover, our SX-ARPES measurements, which are reasonably bulk sensitive, reveal the three-dimensional linearly dispersive bulk Weyl cones and Weyl nodes. Furthermore, by combining the low-photon-energy and SX-ARPES data, we show that the locations of the projected bulk Weyl nodes correspond to the terminations of the Fermi arcs within our experimental resolution. These systematic measurements demonstrate TaAs as a Weyl semimetal.

The material system and theoretical considerations

Tantalum arsenide, TaAs, is a semimetallic material that crystalizes in a body-centered tetragonal lattice system (Fig. 1A) (*23*). The lattice constants are $a = 3.437 \text{ Å}$ and $c = 11.656 \text{ Å}$, and the space group is $I4_1md$ (#109, $C_{4v}$), as consistently reported in previous structural studies (*23-25*). The crystal consists of interpenetrating Ta and As sub-lattices, where the two sub-lattices are shifted by $(\frac{a}{2}, \frac{a}{2}, \delta)$, $\delta \approx \frac{c}{12}$. Our diffraction data matches well with the lattice parameters and the space group $I4_1md$ (*26*). The scanning tunneling microscopic (STM) topography (Fig. 1B) clearly resolves the (001) square lattice without any obvious defect. From the topography, we obtain a lattice constant $a = 3.45 \text{ Å}$. Electrical transport measurements on TaAs confirmed its semimetallic transport properties and reported negative magnetoresistance suggesting the anomalies due to Weyl fermions (*23*).

We discuss the essential aspects of the theoretically calculated bulk band structure (*9, 10*) that predicts TaAs as a Weyl semimetal candidate. Without spin-orbit coupling, calculations (*9, 10*) show that the conduction and valence bands interpenetrate (dip into) each other to form four 1D line nodes (closed loops) located on the $k_x$ and $k_y$ planes (shaded blue in Figs. 1, C and E). Upon the inclusion of spin-orbit coupling, each line node loop is gapped out and shrinks into six Weyl nodes that are away from the $k_x = 0$ and $k_y = 0$ mirror planes (Fig. 1E, small filled circles). In our calculation, in total there are 24 bulk Weyl cones (*9*, *10*), all of which are linearly dispersive and are associated with a single chiral charge of $\pm 1$ (Fig. 1E). We denote the 8 Weyl nodes that are located on the brown plane ($k_z = \frac{2\pi}{c}$) as W1 and the other 16 nodes that are away from this plane as W2. At the (001) surface BZ (Fig. 1F), the 8 W1 Weyl nodes are projected in the vicinity of the surface BZ edges, $\overline{X}$ and $\overline{Y}$. More interestingly, pairs of W2 Weyl nodes with the same chiral charge are projected onto the same point on the surface BZ. Therefore, in total there are 8 projected W2 Weyl nodes with a projected chiral charge of $\pm 2$, which are located near the midpoints of the $\overline{\Gamma} - \overline{X}$ and the $\overline{\Gamma} - \overline{Y}$ lines. Because the $\pm 2$ chiral charge is a projected value, the Weyl cone is still linear (*9*). The number of Fermi arcs terminating on a projected Weyl node must equal its projected chiral charge. Therefore, in TaAs, two Fermi arc surface states must terminate on each projected W2 Weyl node.

Surface electronic structure of TaAs

We carried out low-photon-energy ARPES measurements to explore surface electronic structure of TaAs. Figure 1H presents an overview of the (001) Fermi surface map. We observe three types of dominant features, namely a crescent-shaped feature in the vicinity of the midpoint of each $\overline{\Gamma} - \overline{X}$ or $\overline{\Gamma} - \overline{Y}$ line, a bowtie-like feature centered at the $\overline{X}$ point, and an extended feature centered at the $\overline{Y}$ point. We find that the Fermi surface and the constant energy contours at shallow binding energies (Fig. 2A) violate the $C_4$ symmetry, considering the features at $\overline{X}$ and $\overline{Y}$ points. In the crystal structure of TaAs, where the rotational symmetry is implemented as a screw axis that sends the crystal back into itself after a $C_4$ rotation and a translation by $\frac{c}{2}$ along the rotation axis, such an asymmetry is expected in calculation. The crystallinity of (001) surface in fact breaks the rotational symmetry. We now focus on the crescent-shaped features. Their peculiar shape suggests the existence of two arcs and their termination points in $k$-space seem to coincide with the surface projection of the W2 Weyl nodes. Because the crescent feature consists of two non-closed curves, it can either arise from two Fermi arcs or a closed contour, however, the decisive property that clearly distinguishes one case from the other is the way in which the constant energy contour evolves as a function of energy. As shown in Fig. 2F, in order for the crescent feature to be Fermi arcs, the two non-closed curves have to move (disperse) in the same direction as one varies the energy (*26*). We now provide ARPES data to show that the crescent features in TaAs indeed exhibit this "co-propagating" property. To do so, we single out a crescent feature as shown in Figs. 2, B and E and show the band dispersions at representative momentum space cuts, Cut I and Cut II, as defined in Fig. 2E. The corresponding $E-k$ dispersions are shown in Figs. 2, C and D. The evolution (dispersive "movement") of the bands as a function of binding energy can be clearly read from the slope of the bands in the dispersion maps, and is indicated in Fig. 2E by the white arrows. It can be seen that the evolution of the two non-closed curves are consistent with the co-propagating property. In order to further visualize the evolution of the constant energy contour throughout $k_x, k_y$ space, we use surface state constant energy contours at two slightly different binding energies, namely $E_B = 0 = E_F$ and $E_B = 20$ meV. Figure 2G shows the difference between these two constant energy contours, namely $\Delta I(k_x,k_y) = I(E_B = 20 \text{ meV}, k_x, k_y) - I(E_B = 0 \text{ meV}, k_x, k_y)$, where $I$ is the ARPES intensity. The $k$-space regions in Fig. 2G that have negative spectral weight (red) correspond to the constant energy contour at $E_B = 0$ meV, whereas those regions with positive spectral weight (blue) corresponds to the contour at $E_B = 20$ meV. Thus one can visualize the two contours in a single $k_x, k_y$ map. The alternating "red - blue - red - blue" sequence for each crescent feature in Fig. 2G shows the co-propagating property, consistent with Fig. 2F. Furthermore, we note that there are two crescent features, one located near the $k_x = 0$ axis and the other near the $k_y = 0$ axis, in Fig. 2G. The fact that we observe the co-propagating property for two independent crescent features which are $90°$ rotated with respect to each other further shows that this observation is not due to artifacts, such as a $k$ misalignment while performing the subtraction. The above systematic data reveal the existence of Fermi arcs on the (001) surface of TaAs. Just like one can identify a crystal as a topological insulator by observing an odd number of Dirac cone surface states, we emphasize that our data here are sufficient to identify TaAs as a Weyl semimetal because of bulk-boundary correspondence in topology.

Theoretically, the co-propagating property of the Fermi arcs is unique to Weyl semimetals because it arises from the nonzero chiral charge of the projected bulk Weyl nodes (*26*), which in this case is $\pm 2$. Therefore, this property distinguishes the crescent Fermi arcs not only from any closed contour but also from the double Fermi arcs in Dirac semimetals (*27*, *28*) because the bulk Dirac nodes do not carry any net chiral charges (*26*). After observing the surface electronic structure containing Fermi arcs in our ARPES data, we are able to slightly tune the free parameters of our surface calculation and obtain a calculated surface Fermi surface that reproduces and explains our ARPES data (Fig. 1G). This serves as an important cross-check that our data and interpretation are self consistent. Specifically, our surface calculation indeed also reveals the crescent Fermi arcs that connect the projected W2 Weyl nodes near the midpoints of each $\bar{\Gamma}-\bar{X}$ or $\bar{\Gamma}-\bar{Y}$ line (Fig. 1G). In addition, our calculation shows the bowtie surface states centered at the $\bar{X}$ point, also consistent with our ARPES data. According to our calculation, these bowtie surface states are in fact Fermi arcs (*26*) associated with the W1 Weyl nodes near the BZ boundaries. However, our ARPES data cannot resolve the arc character since the W1 Weyl nodes are too close to each other in momentum space compared to the experimental resolution. Additionally, we note that the agreement between the ARPES data and the surface calculation upon the contour at the $\bar{Y}$ point can be further improved by fine-optimizing the surface parameters. In order to establish the topology, it is not necessary for the data to have a perfect correspondence with the details of calculation because some changes in the choice of the surface potential allowed by the free parameters do not change the topology of the materials, as is the case in topological insulators (*21*, *22*). In principle, Fermi arcs can coexist with additional closed contours in a Weyl semimetal (*6*, *9*), just as Dirac cones can coexist with additional trivial surface states in a topological insulator (*21*, *22*). Particularly, establishing one set of Weyl Fermi arcs is sufficient to prove a Weyl semimetal (*6*). This is achieved by observing the crescent Fermi arcs as we show here by our ARPES data in Fig. 2, which is further consistent with our surface calculations.

Bulk measurements

We now present bulk-sensitive SX-ARPES (*29*) data, which reveal the existence of bulk Weyl cones and Weyl nodes. This serves as an independent proof of the Weyl semimetal state in TaAs. Figure 3B shows the SX-ARPES measured $k_x - k_z$ Fermi surface at $k_y = 0$ (note that none of the Weyl nodes are located on the $k_y = 0$ plane). We emphasize that the clear dispersion along the

$k_z$ direction (Fig. 3B) firmly shows that our SX-ARPES predominantly images the bulk bands. SX-ARPES boosts the bulk-surface contrast in favor of the bulk band structure, which can be further tested by measuring the band dispersion along the $k_z$ axis in the SX-ARPES setting. This is confirmed by the agreement between the ARPES data (Fig. 3B) and the corresponding bulk band calculation (Fig. 3A). We now choose an incident photon energy (i.e. a $k_z$ value) that corresponds to the $k$-space location of W2 Weyl nodes and map the corresponding $k_x - k_y$ Fermi surface. As shown in Fig. 3C, the Fermi points that are located away from the $k_x$ or $k_y$ axes are the W2 Weyl nodes. In Fig. 3D, we clearly observe two linearly dispersive cones that correspond to the two nearby W2 Weyl nodes along Cut 1. The $k$-space separation between the two W2 Weyl nodes is measured to be $0.08 \text{ Å}^{-1}$, which is consistent with both the bulk calculation and the separation of the two terminations of the crescent Fermi arcs measured in Fig. 2. The linear dispersion along the out-of-plane direction for the W2 Weyl nodes is shown by our data in Fig. 3E. Additionally, we also observe the W1 Weyl cones in Figs. 3G-I. Notably, our data shows that the energy of the bulk W1 Weyl nodes is lower than that of the bulk W2 Weyl nodes, which agrees well with our calculation shown in Fig. 3J and an independent modeling of the bulk transport data on TaAs (*23*).

In general, in a spin-orbit coupled bulk crystal, point-like linear band crossings can either be Weyl cones or Dirac cones. Because the observed bulk cones in Figs. 3C,D are located neither at Kramers' points nor on a rotational axis, they cannot be identified as bulk Dirac cones and have to be Weyl cones according to topological theories (*6*, *28*). Therefore, our SX-ARPES data alone, proves the existence of bulk Weyl nodes. The agreement between the SX-ARPES data and our bulk calculation, which only requires the crystal structure and the lattice constants as inputs, provides further cross-check.

Bulk-surface correspondence

Finally, we show that the $k$-space locations of the surface Fermi arc terminations match with the projection of the bulk Weyl nodes on the surface BZ. We superimpose the SX-ARPES measured bulk Fermi surface containing W2 Weyl nodes (Fig. 3C) onto the low-photon-energy ARPES Fermi surface containing the surface Fermi arcs (Fig. 2A) to-scale. From Fig. 4A we see that all the arc terminations and projected Weyl nodes match with each other within the $k$-space region that is covered in our measurements. To establish this point quantitatively, in Fig. 4C, we show the zoomed-in map near the crescent Fermi arc terminations, from which we obtain the $k$-space location of the terminations to be at $\bar{k}_{\text{arc}} = (0.04 \pm 0.01 \text{Å}^{-1}, 0.51 \pm 0.01 \text{Å}^{-1})$. Fig. 4D shows the zoomed-in map of two nearby W2 Weyl nodes, from which we obtain the $k$-space location of the W2 Weyl nodes to be at $\bar{k}_{\text{W2}} = (0.04 \pm 0.015 \text{Å}^{-1}, 0.53 \pm 0.015 \text{Å}^{-1})$. In our bulk calculation, the $k$-space location of the W2 Weyl nodes is found to be at $(0.035 \text{Å}^{-1}, 0.518 \text{Å}^{-1})$. Since the SX-ARPES bulk data and the low-photon-energy ARPES surface data are completely independent measurements using two different beamlines, the fact that they match well provides another piece of evidence of the topological nature (the surface-bulk correspondence) of the Weyl semimetal state in TaAs. In Figs. S5-7, we further show that the bulk Weyl cones can also be observed in our low-photon-energy ARPES data, although their spectral weight is much lower than the surface state intensities that dominate the data. Our demonstration of the Weyl fermion semimetal state in TaAs paves the way for the realization of many fascinating topological quantum phenomena.

**Acknowledgments:** Work at Princeton University and Princeton-led synchrotron-based ARPES measurements were supported by the Gordon and Betty Moore Foundations EPiQS Initiative through Grant GBMF4547 (Hasan). First-principles band structure calculations at National University of Singapore were supported by the National Research Foundation, Prime Minister's Office, Singapore under its NRF fellowship (NRF Award No. NRF-NRFF2013-03). Single crystal growth was supported by National Basic Research Program of China (Grant Nos. 2013CB921901 and 2014CB239302) and characterization by U.S. DOE DE-FG-02-05ER46200. F.C.C acknowledges the support provided by MOST-Taiwan under project number 102-2119-M-002-004. We gratefully acknowledge J. D. Denlinger, S. K. Mo, A. V. Fedorov, M. Hashimoto, M. Hoesch, T. Kim, and V. N. Strocov for their beamline assistance at the Advanced Light Source, the Stanford Synchrotron Radiation Lightsource, the Diamond Light Source, and the Swiss Light Source. We also thank D. Huse, I. Klebanov, T. Neupert, A. Polyakov, P. Steinhardt, H. Verlinde, and A. Vishwanath for discussions. We thank Tay-Rong Chang for help on band structure calculations. R.S. and H.L. acknowledge visiting scientist support from Princeton University.

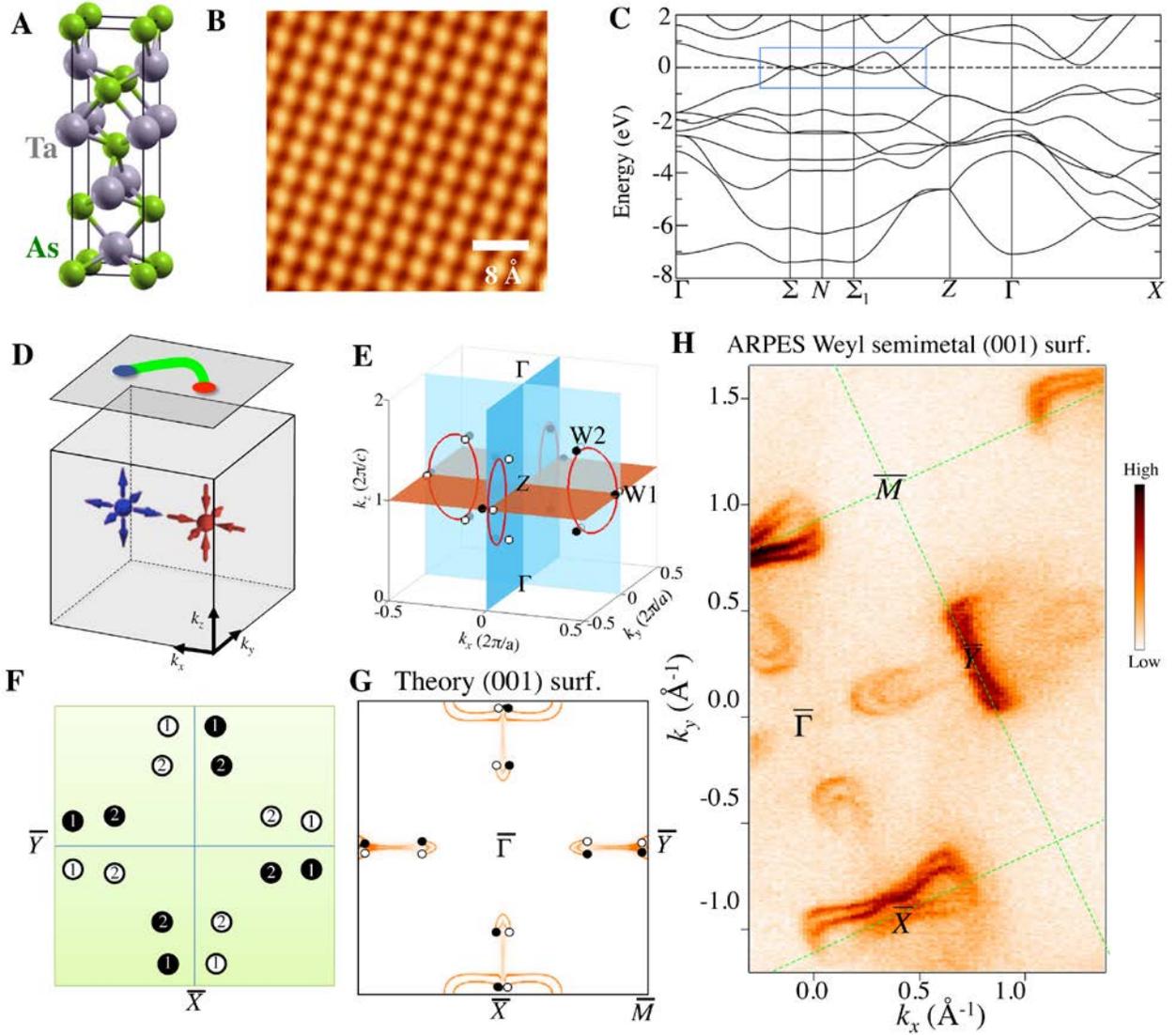

**Fig. 1. Crystal structure and electronic structure of TaAs.** (**A**) Body-centred tetragonal structure of TaAs, shown as stacked Ta and As layers. The lattice of TaAs does not have space inversion symmetry. (**B**) STM topographic image of TaAs's (001) surface taken at the bias voltage -300 mV, revealing the surface lattice constant. (**C**) First-principles band structure calculations of TaAs without spin-orbit coupling. The blue box highlights the locations where bulk bands touch in the BZ. (**D**) Illustration of the simplest Weyl semimetal state that has two

single Weyl nodes with the opposite ($\pm 1$) chiral charges in the bulk. (**E**) In the absence of spin-orbit coupling, there are two line nodes on the $k_x$ mirror plane and two line nodes on the $k_y$ mirror plane (red loops). In the presence of spin-orbit coupling, each line node reduces into six Weyl nodes (small black and white circles). Black and white show the opposite chiral charges of the Weyl nodes. (**F**) A schematic (not to-scale) showing the projected Weyl nodes and their projected chiral charges. (**G**) Theoretically calculated band structure (*26*) of the Fermi surface on the (001) surface of TaAs. (**H**) The ARPES measured Fermi surface of the (001) cleaving plane of TaAs. The high symmetry points of the surface BZ are noted.

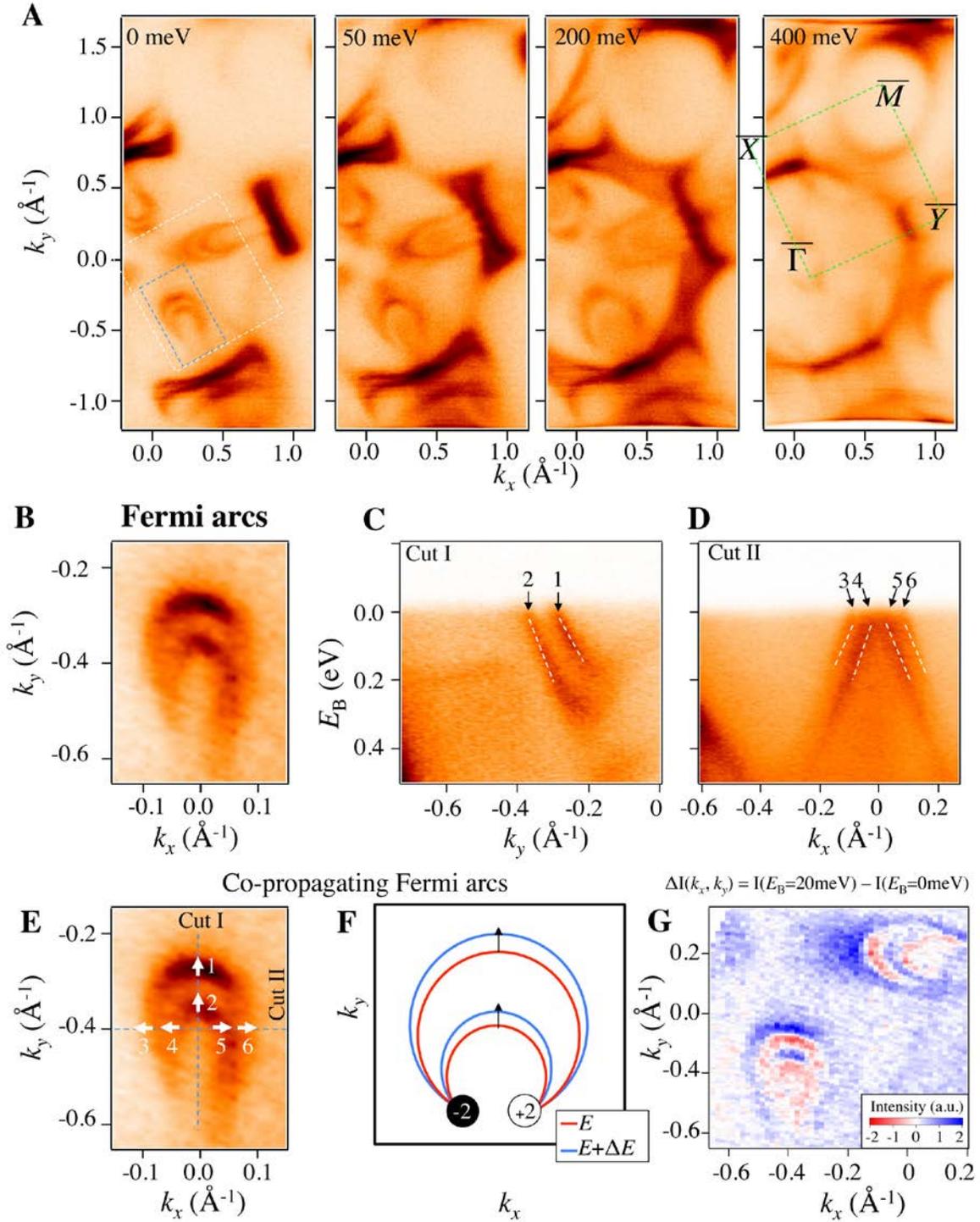

**Fig. 2. Observation of topological Fermi arc surface states on the (001) surface of TaAs.** (**A**) ARPES Fermi surface map and constant binding energy contours measured using incident photon energy of 90 eV. (**B**) High-resolution ARPES Fermi surface map of the crescent Fermi

arcs. The $k$-space range of this map is defined by the blue box in panel A. (**C,D**) Energy dispersion maps along Cuts I and II. (**E**) Same Fermi surface map as in panel B. The dotted lines define the $k$-space direction for Cuts I and II. The numbers 1-6 note the Fermi crossings that are located on Cuts I and II. The white arrows show the evolution of the constant energy contours as one varies the binding energy, which is obtained from the dispersion maps in panels C and D. (**F**) A schematic showing the evolution of the Fermi arcs as a function of energy, which clearly distinguish between two Fermi arcs and a closed contour. (**G**) The difference between the constant energy contours at the binding energy $E_B = 20$ meV and the binding energy $E_B = 0$ meV, from which one can visualize the evolution of the constant energy contours through $k_x - k_y$ space. The range of this map is shown by the white dotted box in panel A.

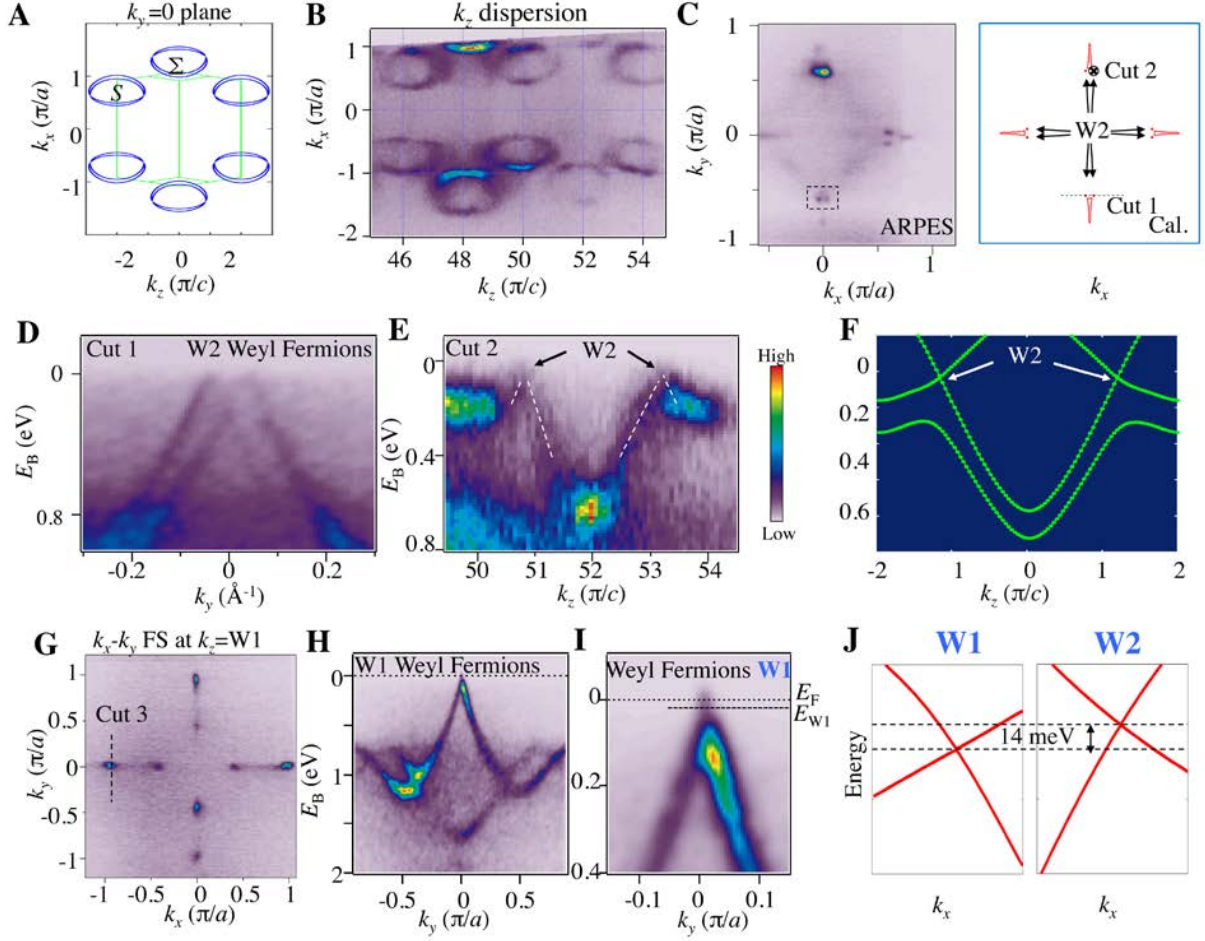

**Fig. 3. Observation of bulk Weyl Fermion cones and Weyl nodes in TaAs.** (**A,B**) First-principles calculated and ARPES measured $k_z - k_x$ Fermi surface maps at $k_y = 0$, respectively. (**C**) ARPES measured and first-principles calculated $k_x - k_y$ Fermi surface maps at the $k_z$ value that corresponds to the W2 Weyl nodes. The dotted line defines the $k$-space cut direction for Cut 1, which goes through two nearby W2 Weyl nodes along the $k_y$ direction. The black cross defines Cut 2, which means that the $k_x, k_y$ values are fixed at the location of a W2 Weyl node and one varies the $k_z$ value. (**D**) ARPES $E - k_y$ dispersion map along the Cut 1 direction, which clearly shows the two linearly dispersive W2 Weyl cones. (**E**) ARPES $E - k_z$ dispersion map along the Cut 2 direction, showing that the W2 Weyl cone also disperses linearly along the out-

of-plane $k_z$ direction. (**F**) First-principles calculated $E - k_z$ dispersion that corresponds to the Cut 2 shown in panel E. (**G**) ARPES measured $k_x - k_y$ Fermi surface maps at the $k_z$ value that corresponds to the W1 Weyl nodes. The dotted line defines the $k$-space cut direction for Cut 3, which goes through the W1 Weyl nodes along the $k_y$ direction. (**H,I**) ARPES $E - k_y$ dispersion map and its zoomed-in version along the Cut 3 direction, revealing the linearly dispersive W1 Weyl cone. (**J**) First-principles calculation shows a $14\ \text{meV}$ energy difference between the W1 and W2 Weyl nodes.

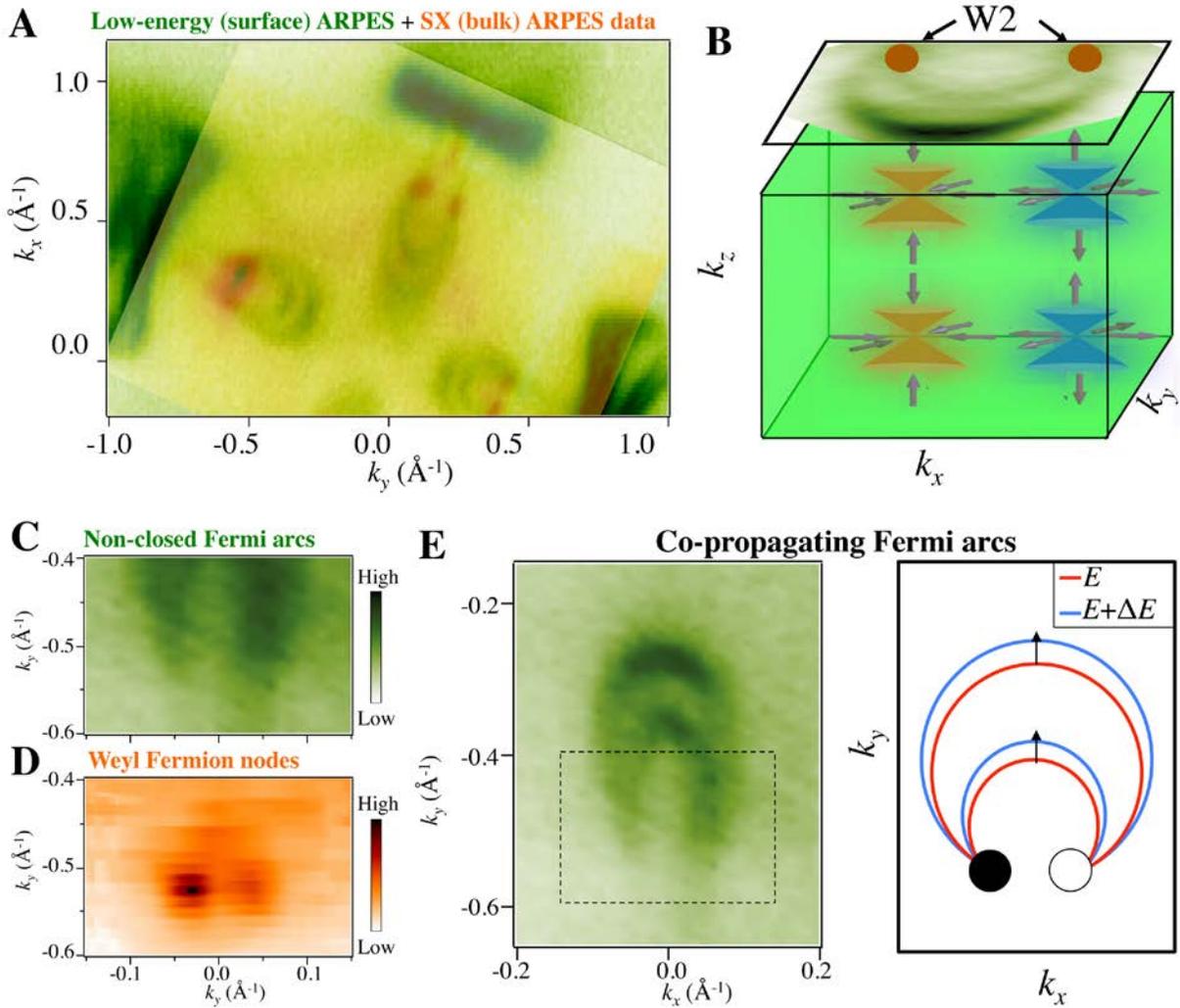

**Fig. 4. Surface-bulk correspondence and the topologically nontrivial state in TaAs.** (**A**) Low-photon-energy ARPES Fermi surface map ($h\upsilon = 90$ eV) from Fig. 2A, with the SX-ARPES map ($h\upsilon = 650$ eV) from Fig. 3C overlaid on top of it to-scale, showing that the locations of the projected bulk Weyl nodes correspond to the terminations of the surface Fermi arcs. (**B**) The bottom shows a rectangular tube in the bulk BZ that encloses four W2 Weyl nodes. These four W2 Weyl nodes project onto two points at the (001) surface BZ with projected chiral charges of $\pm 2$, shown by the brown circles. The top surface shows the ARPES measured crescent surface Fermi arcs that connect these two projected Weyl nodes. (**C**) Surface state Fermi surface map at the $k$-space region corresponding to the terminations of the crescent Fermi arcs. The $k$-space region is defined by the black dotted box in panel E. (**D**) Bulk Fermi surface map at the $k$-space region corresponding to the W2 Weyl nodes. The $k$-space region is defined by the black dotted box in Fig. 3C. (**E**) ARPES and schematic of the crescent-shaped co-propagating Fermi arcs.